\documentclass[epj]{webofc}
\usepackage[varg]{txfonts}   
\usepackage{graphicx}
\usepackage{amsmath}
\usepackage{amssymb}
\usepackage{pifont}
\usepackage{footnote}

\usepackage{color}
\usepackage{natbib}
\usepackage{hyperref} 
\usepackage{ulem}
\usepackage{graphicx}
\usepackage{wasysym}
\usepackage{cleveref}
\RequirePackage{lineno}

\usepackage[utf8]{inputenc}

\usepackage{xcolor,colortbl}
\usepackage{pifont}
\usepackage{footnote}

\def \beq{\begin{equation}}
\def \eeq{\end{equation}}
\def \beqa{\begin{eqnarray}}
\def \eeqa{\end{eqnarray}}
\def \la{\langle}
\def \ra{\rangle}

\woctitle{}
\begin{document}
\title{Studies of three-particle correlations and reaction-plane correlators from STAR}

\author{Prithwish Tribedy (for the STAR Collaboration) \inst{1}\fnsep\thanks{\email{ptribedy@bnl.gov}}}

\institute{Physics Department, Brookhaven National Laboratory, Upton, NY 11973, USA}

\abstract{%
  We present STAR measurements of various harmonics of three-particle correlations in $\sqrt{s_{NN}}=200$ GeV Au+Au collisions at RHIC. The quantity $\langle\cos(m\phi_1+n\phi_2-(m+n)\phi_3)\rangle$ is measured for inclusive charged particles for different harmonics $m$ and $n$ as a function of collision centrality, transverse momentum $p_T$ and relative pseudorapidity $\Delta\eta$. These observables provide detailed information on global event properties like correlations between event planes of different harmonics and are particularly sensitive to the expansion dynamics of the matter produced in the collisions. We compare our measurements to different viscous hydrodynamic models. We argue that these measurements probe the three dimensional structure of the initial state and provide unique ways to constrain the transport parameters involved in hydrodynamic modeling of heavy-ion collisions.
}
\maketitle
\section{Introduction}
\label{intro}


By now it has been established that relativistic heavy-ion collisions produce a strongly correlated Quark Gluon Plasma (sQGP). Recent experimental studies have focused on studying the properties of such sQGP. One of the striking properties of such a phase of matter is that it exhibits nearly perfect fluidity characterized by the smallest viscosity-to-entropy-density ratio $\eta/s$ amongst all known fluids in the nature. Over the past years, combined insights from both experiment through measurements of anisotropic flow coefficients $v_n$ and from theory through viscous hydrodynamic simulations, have made precise extraction of $\eta/s$ possible. However a very intrinsic characteristic about such transport parameter, i.e. its temperature dependence is not yet fully constrained by experimental measurements. A primary goal of the current measurements at RHIC is therefore to go beyond conventional measurements of flow coefficients and provide new observables that can be useful to constrain $\eta/s (T)$.  

In this work we present the measurements of three particle correlations from the STAR experiment using the observable $C_{m,n,m+n} = \la \la \cos(m\phi_1 + n \phi_2 - (m+n) \phi_3)\ra\ra$, where $m,n$ defines the harmonic coefficients and $\phi_{1,2,3}$, the azimuthal angles of three particles~\cite{Bhalerao:2013ina}. The inner and the outer averages are taken over all triplets and events respectively. The observable $C_{m,n,m+n}$ can be approximated as correlations of flow harmonics, $v_n$s, and corresponding event plane angles, $\Psi_n$s, as ${\la v_m v_n v_{m+n} \cos(m \Psi_m + n \Psi_n - (m+n) \Psi_{m+n}) \ra}$. Theoretical studies show that such an observable can probe non-linear hydrodynamic response and therefore become more sensitive to viscosity than individual flow harmonics $v_n$~\cite{Qiu:2012uy, Teaney:2010vd, Teaney:2012ke, Teaney:2013dta, Bhalerao:2013ina, Yan:2015jma, Qian:2016fpi, Qian:2016pau,McDonald:2016vlt,Betz:2016ayq}. Better sensitivity to viscous effects can be very useful towards more precise extraction of different transport parameters and possibly their temperature dependence by comparison to hydrodynamic simulations~\cite{Niemi:2015qia, Denicol:2015nhu}. 
Measurements of event plane and flow harmonic correlations have been performed at LHC by ATLAS collaboration~\cite{Aad:2014fla} and recently by ALICE collaboration\cite{ALICE:2016kpq}. However measurements at a single energy is not sufficient to constrain $\eta/s~(T)$. LHC measurements are sensitive to the $\eta/s$ at higher temperatures, meanwhile full constraint on $\eta/s~(T)$ can only be achieved with complementary measurements of $C_{m,n,m+n}$ at RHIC~\cite{Niemi:2011ix, Gale:2012rq, Niemi:2015qia, Denicol:2015nhu}. In fact, measurements over the entire range of energy available under the Beam Energy Scan (BES) program at RHIC will be most preferable in this context. Measurements at RHIC have additional advantages. Since the beam rapidity is smaller one expects stronger variation of initial geometry, fluctuations, energy density, temperature, baryon density etc. over a relatively smaller window of rapidity as compared to LHC. In this context, measurements of $C_{m,n,m+n}$ on the pseudorapidity separation between particles may allow us to study the breaking of longitudinal invariance, three dimension structure of the initial state~\cite{Teaney:2010vd, Bozek:2010vz,Jia:2014ysa,Pang:2015zrq,Schenke:2016ksl} over relatively smaller widow of acceptance available for measurements at RHIC than LHC. In addition to constraining initial state and transport parameters, the charge dependence of three particle correlation can be used to search for the signals of the chiral magnetic effect (CME)~\cite{Abelev:2009ac, Abelev:2009ad, Adamczyk:2013hsi, Voloshin:2004vk}. In this work we will not study such charge dependence, however, we expect that the results presented here for inclusive charged particles will provide important baseline for the CME measurements.

\section{Experiment and analysis}
We analyze the data on Au+Au collisions at $\sqrt{s_{NN}}=200$ GeV collected by the STAR detector~\cite{Ackermann2003624} during 2011 year running of RHIC. For the measurements of $C_{m,n,m+n}$ we use charged particles within the pseudorapidity range of $|\eta|\!<\!1$ and transverse momentum of $p_T\!>\!0.2$ GeV/$c$ detected by the Time Projection Chamber (TPC), the primary tracking systems of STAR situated inside a 0.5 Tesla solenoidal magnetic field~\cite{Anderson2003659}. We use algebra based on Q-vectors and in order to account for imperfections in the detector acceptance we apply track-by-track weights~\cite{Bilandzic:2010jr, Bilandzic:2013kga}. We also apply momentum dependent tracking efficiency. In such estimation, we correct for the track-merging artifacts by measuring the relative pseudo rapidity separation between any two tracks and correcting for missing pairs apparent at $\Delta\eta \approx 0$. We estimate systematic uncertainties in our measurements by analyzing datasets of different time periods, from different years, with different tracking algorithms, with different efficiency estimates, by varying z-vertex position of the collision, and by varying track selection criteria. In addition we also quantify the effects of short-range quantum and Coulomb correlations in the systematic uncertainties by studying $\Delta\eta$ dependence of $C_{m,n,m+n}$. Finally for data-model comparison we estimate the number of participant nucleons $N_{\rm part}$ using a Monte-Carlo Glauber model for different centrality intervals ($0-5\%, 5-10\%, 10-20\%, ..., 70-80\%$) used in this analysis~\cite{Abelev:2008ab,Miller:2007ri}. For selection of such centrality bins we use the distribution of minimum bias uncorrected multiplicity of charged particles in the pseudorapidity region $|\eta|<0.5$ measured by the TPC.

\section{Results and discussion}
 \begin{figure}[t]
\includegraphics[width=\textwidth]{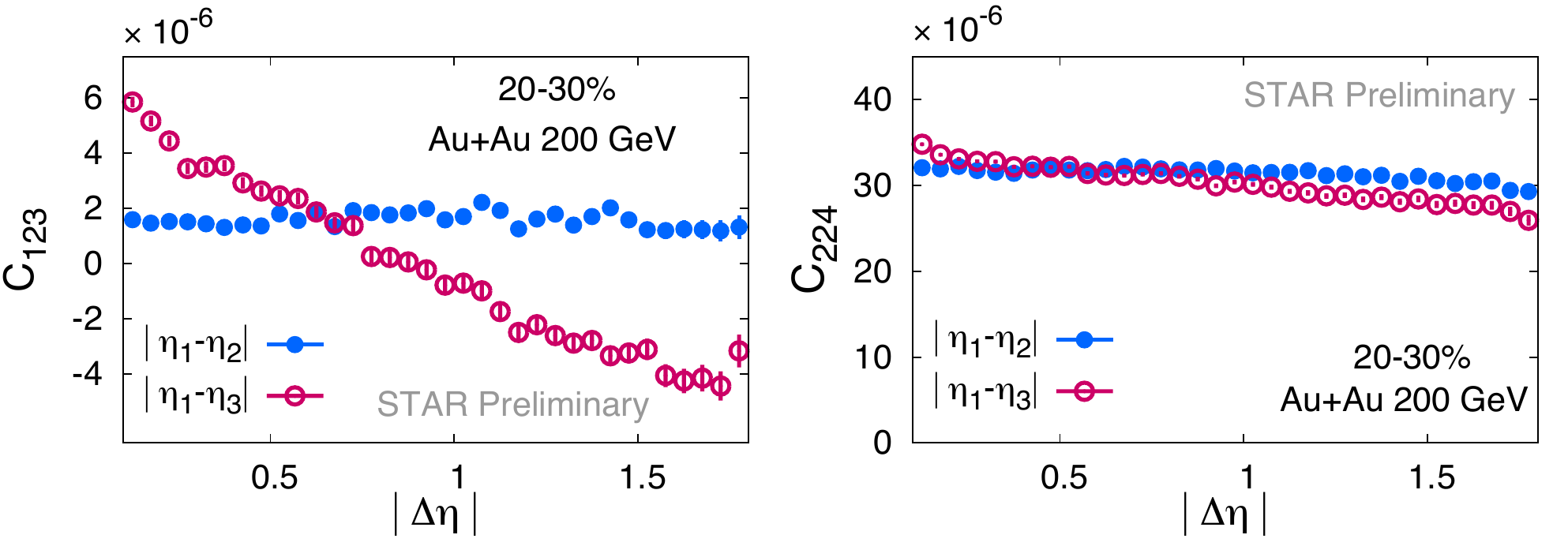}
\caption{\small \label{detdist} Relative pseudo rapidity dependence of the three particle correlator $C_{1,2,3}$ and $C_{2,2,4}$.}
\end{figure}
In this conference proceedings we present results for the correlators $C_{1,1,2}, C_{1,2,3}, C_{2,2,4}, C_{2,3,5}$. We first present differential measurements such as $\Delta\eta$ and $p_T$ dependence of these correlators, the goal of such study is to understand how different physical scenarios effect these observables. We later on present integrated measurements i.e. the centrality dependence  of $C_{m,n,m+n}$ and make comparisons to viscous hydrodynamic model calculations with different assumptions of $\eta/s (T)$. 

 \subsection{$\Delta\eta$ dependence}
 \vspace{-5pt}
In Fig.~\ref{detdist} we show the $\Delta\eta$ dependence of $C_{1,2,3}$ and $C_{2,2,4}$. A similar measurement for $C_{1,1,2}$ was previously presented by STAR in Ref~\cite{Abelev:2009ad}~\footnote{The $\Delta\eta$ dependence for all other harmonics of  $C_{m,n,m+n}$ will be presented in a future publication.}. One can clearly see in Fig.~\ref{detdist}(left) that $C_{1,2,3}$ correlator shows a very strong dependence on $\Delta\eta_{1,3}$ (i.e. between the first and third order harmonics) but a very weak dependence on $\Delta\eta_{1,2}$ (i.e. between first and second order harmonics). We omit the curve for the variation of $C_{1,2,3}$ with $\Delta\eta_{2,3}$ for clarity which looks very similar to the curve shown for $\Delta\eta_{1,2}$. A strong dependence of $C_{1,1,2}$ correlator for $\Delta\eta_{1,2}$ (i.e. between two first order harmonics) was also observed in the previous STAR measurement in Ref~\cite{Abelev:2009ad}. In contrast, the similar measurement shown in Fig.\ref{detdist} (right) for the correlator $C_{2,2,4}$ with two possible combinations of $\Delta\eta$ shows a much weaker dependence compared to its absolute magnitude. These observations indicate a very specific pattern for three particle correlations. The relative rapidity dependence between either ``first-first'' or ``first-third'' harmonics show strong variations and even change of sign, whereas between second and any other harmonics the correlations show much weaker variation in relative rapidity. 

Variations of $C_{m,n,m+n}$ with $\Delta\eta$ can come from hydrodynamic response to the three-dimensional structure of initial state~\cite{Teaney:2010vd, Bozek:2010vz,Jia:2014ysa,Pang:2015zrq,Schenke:2016ksl}. They can also arise from artifacts such as short-range correlations, non-flow and resonance decays~\cite{Abelev:2009ac}, etc., that give rise to two-particle correlations that are correlated to an event plane (determined by the third particle) and do not vanish after averaging over many events.  
However, if such a variation persist up to large $\Delta\eta$, e.g. as shown in Fig.\ref{detdist} for $C_{1,2,3}$ vs $\Delta\eta_{1,3}$, they can not be driven by short range correlations. 
In a flow scenario, strong variation in $\Delta\eta$ can come from de-correlation in initial state geometry, e.g. driven by a breaking of longitudinal invariance through a forward-backward rapidity dependence of harmonic planes particularly between $\Psi_1$ and $\Psi_3$~\cite{Teaney:2010vd}. In case of $\Psi_2$, one do not expect strong variation with rapidity due to geometry of collisions. 
In a non-flow scenario, in case of $C_{1,2,3}$ one possible source of $\Delta\eta_{1,3}$ dependence could be momentum conservation that leads to back-to-back correlations between two particles from jets that are correlated to second order event plane. We discuss such scenario in the next section. 
 \subsection{$p_T$ dependence}

\begin{figure}[t]
\includegraphics[width=\textwidth]{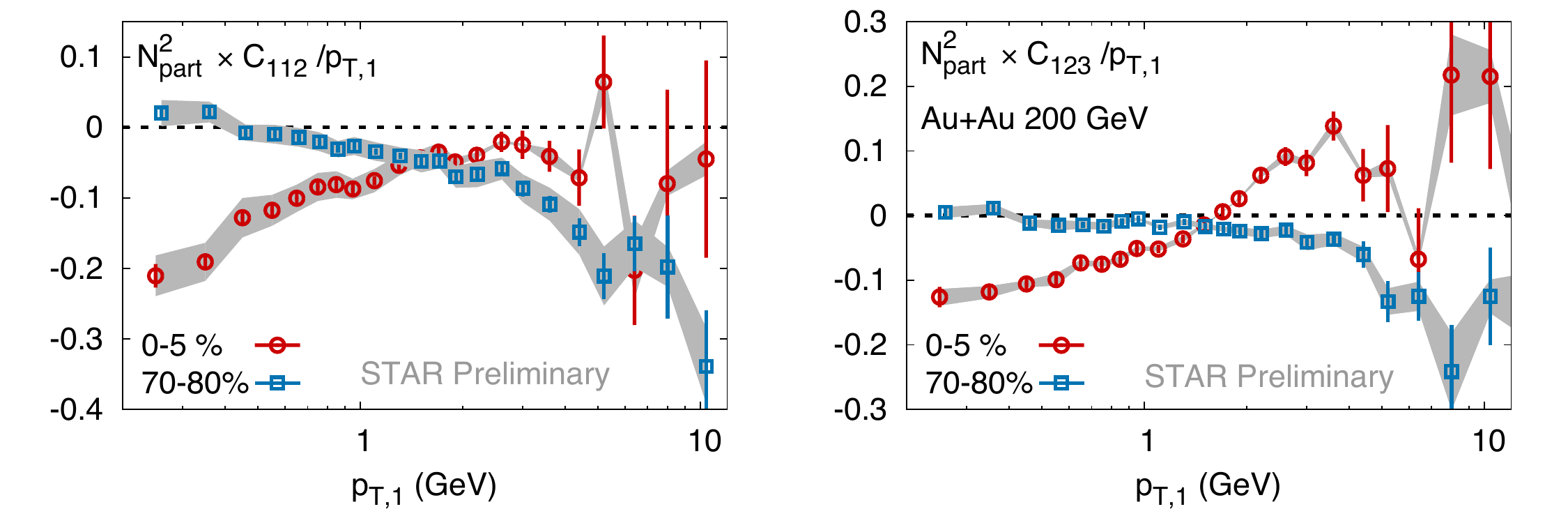}
\caption{\small \label{ptdist} Transverse momentum dependence of the three particle correlator $C_{1,1,2}$ and $C_{1,2,3}$.}
\end{figure}

The effect of momenta conservation is expected to be dominant at higher transverse momentum and for low multiplicity events. Therefore, measurements performed in peripheral events can be a good baseline for such studies. In the central events due to large number of particles, quenching of jet-like correlations etc., the effect of momentum conservation will not be dominant.  
It is therefore essential to perform this exercise in both central (e.g. $0\!-\!5\%$) and peripheral (e.g. $70\!-\!80\%$) and contrast the trend seen in data. 
From $\Delta\eta$ dependence of $C_{m,n,m+n}$, as discussed in previous section, we find that the correlators involving first order harmonics can be sensitive to non-flow effects such as momentum conservations from jets etc. In Fig.~\ref{ptdist} we therefore study the variations of the correlators $C_{1,1,2}$ and $C_{1,2,3}$ with the transverse momentum $p_T$ of the particle corresponding to the first order harmonic, i.e. for the first particle as denoted by $p_{T,1}$. In order to remove trivial increase of first order harmonic $v_1$ with transverse momentum and trivial dilution of correlation while going from peripheral to central events we multiply the correlator by a factor of $N_{\rm part}^2/p_{T,1}$. The results for $70\!-\!80\%$ indicates that at high $p_{T,1}$ both $C_{1,1,2}$ and $C_{1,2,3}$ becomes negative. Such trend is consistent with a picture of momentum conservation and can be understood as follows. If a pair of back-to-back particles gets aligned along $\Psi_2$, they will lead to negative values for these correlators since then we have $C_{1,1,2}\approx \cos(\pi)$ and $C_{1,2,3}\approx\cos(\pm3\pi)$. This might explain the decreasing trend for $70\!-\!80\%$ events. However such a scenario can not explain the trend seen $0-5\%$ events where one finds negative signal at small $p_{T,1}$ and nearly zero or positive signal at large $p_{T,1}$. This qualitatively different trend seen in central events can not be due to non-flow correlations from back-to-back pairs. 

Clearly the differential measurements of these correlators can provide better insights of the relative contributions of different sources of correlations that can affect $C_{m,n,m+n}$. Model calculations that include full treatment of three-dimensional initial geometry, fluctuations and different other sources of correlations can improve our understanding in this context~\cite{Longacre:2016xwm}.  

\subsection{Centrality dependence} 

\begin{figure}[t]
\includegraphics[width=\textwidth]{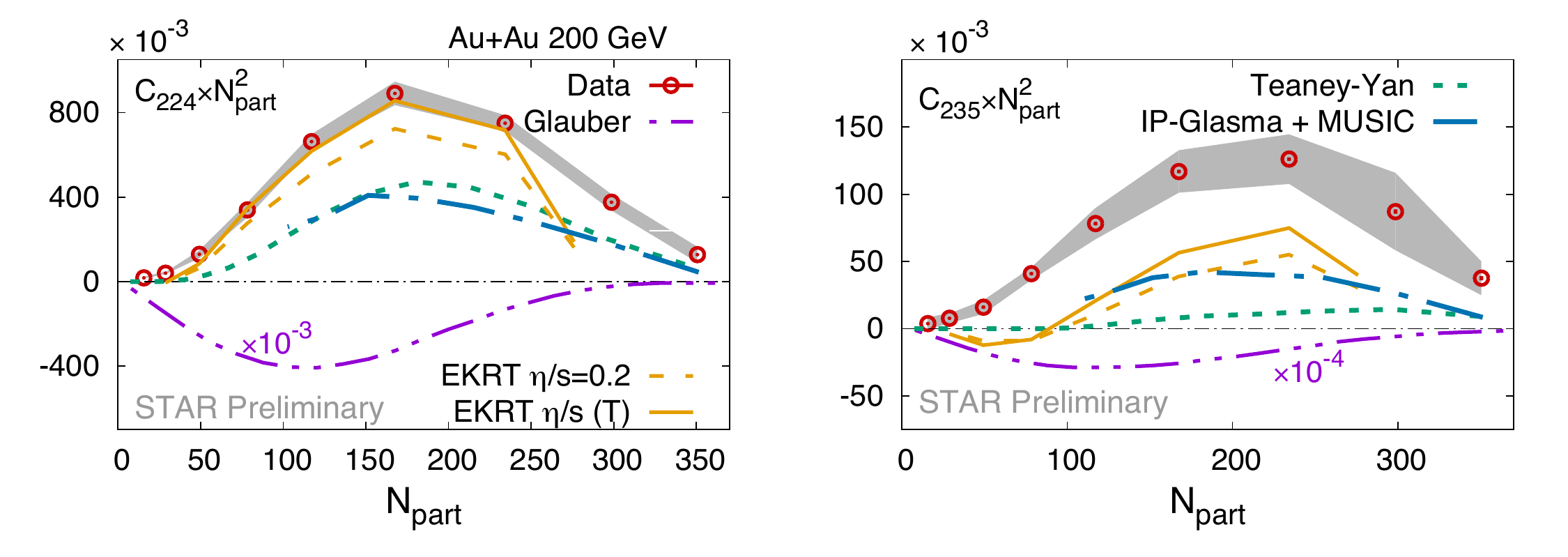}
\caption{\small \label{centdist} Centrality dependence of three particle correlator $C_{1,1,2}$ and $C_{2,2,4}$ compared to different viscous hydrodynamic model calculations.}
\end{figure}

We measure the centrality dependence of $C_{m,n,m+n}$ and compare our results with three different viscous hydrodynamic model calculations. They include 1) hydrodynamic simulations by Teaney and Yan~\cite{Teaney:2010vd,Teaney:2013dta}, 2) the perturbative QCD$+$saturation$+$hydro based ``EKRT" model~\cite{Niemi:2015qia} and 3) hydrodynamic simulations MUSIC~\cite{Schenke:2010nt} with IP-Glasma initial conditions~\cite{Schenke:2012wb}. In addition we also estimate the correlations from initial state geometry using Monte Carlo Glauber model by approximating $C_{m,n,m+n}={\la \varepsilon_m \varepsilon_n \varepsilon_{m+n} \cos(m \Phi_m + n \Phi_n - (m+n) \Phi_{m+n}) \ra}$, where $\varepsilon_n$s and $\Phi_n$s are the initial eccentricities and the participant planes respectively. 
All of these models have been previously constrained by the measurements of $v_n$ and other data on azimuthal correlations from RHIC and LHC, but they do not include longitudinal dependence in the initial state and assume boost invariance. From Fig.\ref{detdist} it is evident that the correlator $C_{2,2,4}$ has the least variation on $\Delta\eta$ and will provide the best opportunity for comparison to boost-invariant hydrodynamic simulations. We therefore present the centrality dependence of the correlator $C_{2,2,4}$ in Fig.\ref{centdist} (left). In Fig.\ref{centdist} (right) we also compare the centrality dependence of $C_{2,3,5}$. In order to scale out the trivial dilution of correlations due to increase of number of pairs while going from peripheral to central events we have multiplied the correlators by $N_{\rm part}^2$. 
 The Glauber model calculations predict that purely initial state correlation of eccentricities and participant planes leads to negative values for all the correlators. Both $C_{2,2,4}$ and $C_{2,3,5}$ being positive in data indicates the dominance of non-linear hydrodynamic response of the medium to initial state geometry. This observation is consistent to the measurement at LHC by the ATLAS collaboration in Ref~\cite{Aad:2014fla}. We however, find that although the qualitative trends predicted by different viscous hydrodynamic simulations are similar to data, some quantitative differences exist. Particularly for $C_{2,2,4}$ one can see that the current precision of the data can very well differentiate between constant and temperature dependent viscosity used in the EKRT simulations. Such comparisons would be key to constrain $\eta/s (T)$. Apart from analysis at $\sqrt{s_{NN}}=200$ GeV, our future studies will be focused on measurements for other lower energies under RHIC Beam Energy Scan program which will provide better constraints of $\eta/s (T)$. 
 
\section*{Summary}
In summary, we have presented the first measurements of three-particle correlations $C_{m,n,m+n}=  \la \la \cos(m\phi_1 + n \phi_2 - (m+n) \phi_3)\ra\ra$ in $\sqrt{s_{NN}}=200$ GeV Au+Au collisions at RHIC. In comparison to conventional flow harmonic measurements these correlators can provide additional information such as de-correlation of event planes driven by three dimensional structure of the initial state and non-linear hydrodynamic response of the medium. When compared to viscous hydrodynamic models these measurements with the precision presented here have the potential to constrain transport parameters and their temperature dependence. 

\section*{Aknowledgement}
This work was supported under Department of Energy Contract No. DE-SC0012704. We thank Li Yan, Risto Paatelainen, Harri Niemi and Gabriel Denicol for providing their model predictions and helpful discussion. 

\bibliography{3pcshort}

\end{document}